\documentclass[%
 aip,
 amsmath,amssymb,
 reprint,%
]{revtex4-1}

\usepackage{graphicx}
\usepackage{dcolumn}
\usepackage{bm}

\usepackage[utf8]{inputenc}
\usepackage[T1]{fontenc}
\usepackage{mathptmx}
\usepackage{todonotes}
\def\be{\begin{equation}}
\def\ee{\end{equation}}

\usepackage{verbatim}

\begin{document}

\preprint{AIP/123-QED}

\title[]{Instability of a smooth shear layer through wave interactions}

\author{J.R. Carpenter}
 \affiliation{Institute of Coastal Research, Helmholtz-Zentrum Geesthacht.}
\author{A. Guha}%
\affiliation{Institute of Coastal Research, Helmholtz-Zentrum Geesthacht.}
\affiliation{Department of Mechanical Engineering, Indian Institute of Technology Kanpur}


\date{\today}

\begin{abstract}
Wave interaction theory can be used as a tool to understand and predict instability in a variety of homogeneous and stratified shear flows.  It is however, most often limited to piecewise-linear profiles of the shear layer background velocity, in which stable vorticity wave modes can be easily identified, and their interaction quantified.  This approach to understanding shear flow instability is extended herein to smooth shear layer profiles.  We describe a method, by which the stable vorticity wave modes can be identified, and show that their interaction results in an excellent description of the stability properties of the smooth shear layer, thus demonstrating the presence of the wave interaction mechanism in smooth shear flows.
\end{abstract}

\maketitle

%


Wave interaction theory (WIT) has been used to provide a physical explanation for many results in the stability of homogeneous and density stratified shear flows \cite{bain1994,carp2013,guha2014wave,book}.  In such a description, an unstable mode results from the interaction of two otherwise freely propagating waves and requires: (i) that the waves are able to adjust each others' phase speed in order to become phase-locked, and (ii) to be configured such that they can cause mutual growth in one another \cite{rede2001,carp2013,guha2014wave,book}.  However, except for a few exceptions \cite{holm1962,carp2010b,iga2013,eyal}, the application of WIT is restricted to idealised piecewise-linear profiles with kinks in the background flow, $U(z)$, where $z$ represents the across-stream distance.  This results in delta function behaviour of background flow vorticity gradients, $U''(z)$, with primes indicating ordinary differentiation with respect to $z$.  Although considerably simplifying the analysis, use of piecewise profiles hampers the applicability of WIT by restricting it to particularly idealised applications.

A major difficulty in applying WIT to more realistic smooth profiles lies in the ambiguity of identifying neutral waves in such flows.  This is due to the fact that the stable normal modes no longer occupy a discrete set of wave modes at a few well-defined phase speeds, $c_n(k)$, with $k$ the horizontal wavenumber, and $n$ a countable index, as they do in piecewise profiles.  Rather, they consist of a continuous $c$-spectrum with a singularity in the form of a vortex sheet located at the critical layer height, $z_c$, where the modal phase speed matches the background flow, i.e., $U(z_c) = c$.  Despite this difficulty in applying WIT to smooth profiles, the stability analysis of piecewise profiles are found to be in qualitative agreement with their smooth counterparts, suggesting that the mechanisms are similar.  In this letter we identify stable wave modes (i.e., vorticity waves) in a smooth shear layer profile, and use these modes to quantify the wave interactions present.  The results lead to an accurate description of the normal mode instability of the shear layer, and demonstrate that we may apply WIT to understand various aspects of the stability of smooth shear flows. 

The stability of two-dimensional perturbations to an inviscid background flow $U(z)$, are governed by Rayleigh's equation \cite{book,draz1982,schmid2001stability}
\begin{equation} \label{eq:rayleigh}
(U-c) (\hat{w}'' - k^2 \hat{w}) - U'' \hat{w} = 0 ,
\end{equation}
for the perturbation vertical velocity given by the normal mode form $w(x,z) = \mathrm{Re}\{ \hat{w}(z)e^{ik(x-ct)} \}$, with $x$ the distance in the direction of the basic flow, and $t$ representing time.  Together with the boundary conditions that $\hat{w}=0$ at $z = z_l, z_u$, Eq. (\ref{eq:rayleigh}) represents an eigen-problem with eigenvalue $c$, and eigenfunction $\hat{w}(z)$, which are functions of the horizontal wavenumber, $k$.  Since the coefficients of Eq. (\ref{eq:rayleigh}) are general functions of $z$, solutions are most often found through numerical computation (see, e.g., Hazel\cite{hazel1972}).  In this letter, we use a second-order finite difference scheme to convert the differential eigenvalue problem in (\ref{eq:rayleigh}) to a matrix eigenvalue problem, which is solved using standard software, as described in Smyth and Carpenter\citep{book}.  For $U(z)$ profiles that do not have an inflection point, all $c$ are necessarily real, and the solution to Eq. (\ref{eq:rayleigh}) is known to be composed of a continuous spectrum of $c$ values with a slope discontinuity of the corresponding eigenfunction at the critical height \cite{draz1982,drazin2002introduction}.  In a numerical solution, the continuous spectrum expresses itself through eigenvalues of $c$ located at the speed of the background flow matching the $z$-grid levels, i.e., $c_j = U(z_j)$ for grid levels at $z_j$.  

\begin{figure}
\begin{center}
\includegraphics[width=0.48\textwidth]{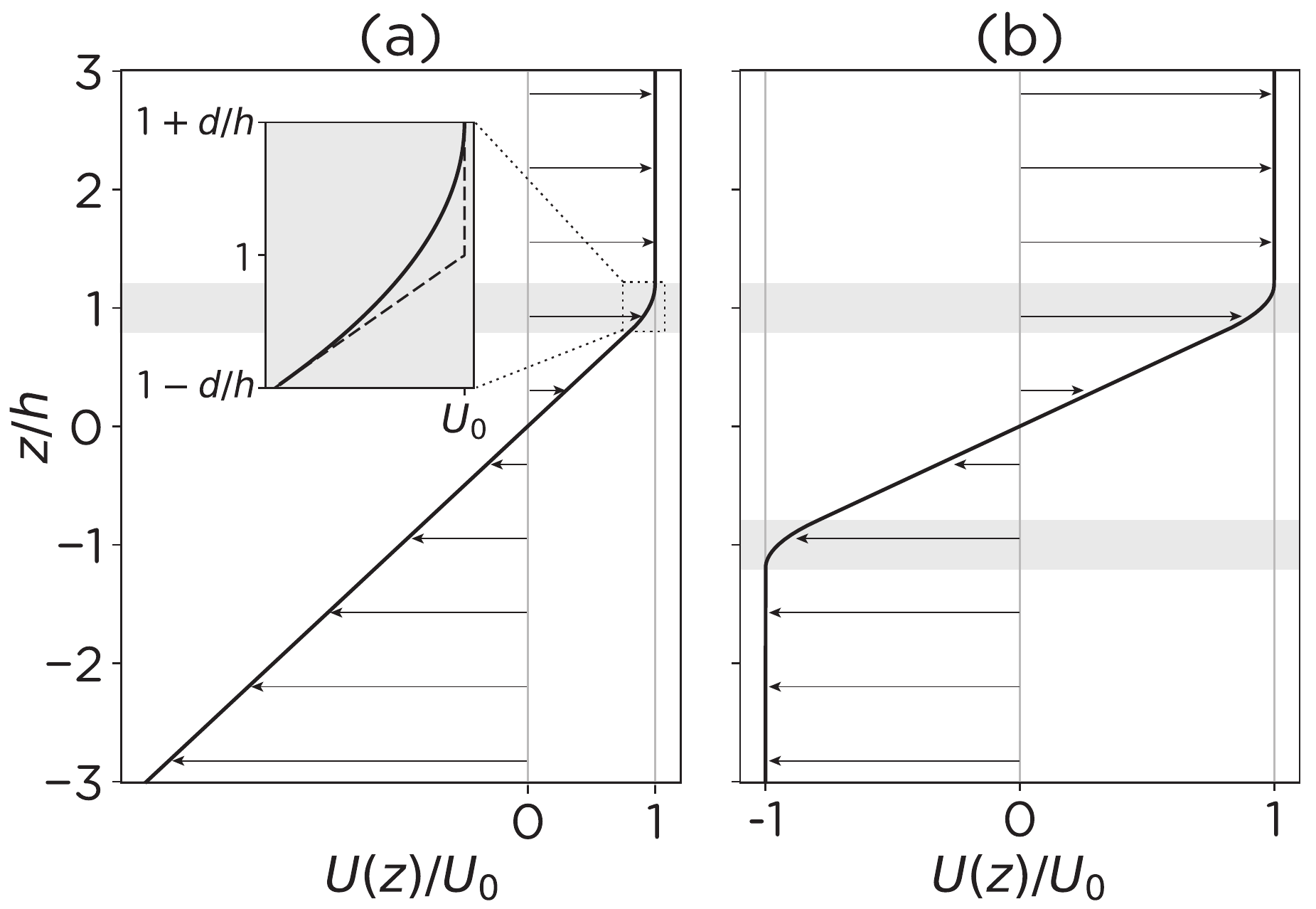} 
\caption{Profiles of (a) a smooth vorticity interface exhibiting stable vorticity waves, and (b) an unstable smooth shear layer constructed as a composite of two smooth vorticity interfaces from the profile in (a).  Regions in which $U''(z) \neq 0$ are highlighted in grey, and each have a thickness of $d = 0.2h$, with vertical boundaries located at $z=\pm 3h$.  Piecewise-linear profiles differ from the smooth profiles only in the grey regions, and are shown as an inset in (a), with a similar definition for the piecewise-linear shear layer.  }
\label{f:profiles}
\end{center}
\end{figure}

\begin{figure*}
\begin{center}
\includegraphics[width=0.95\textwidth]{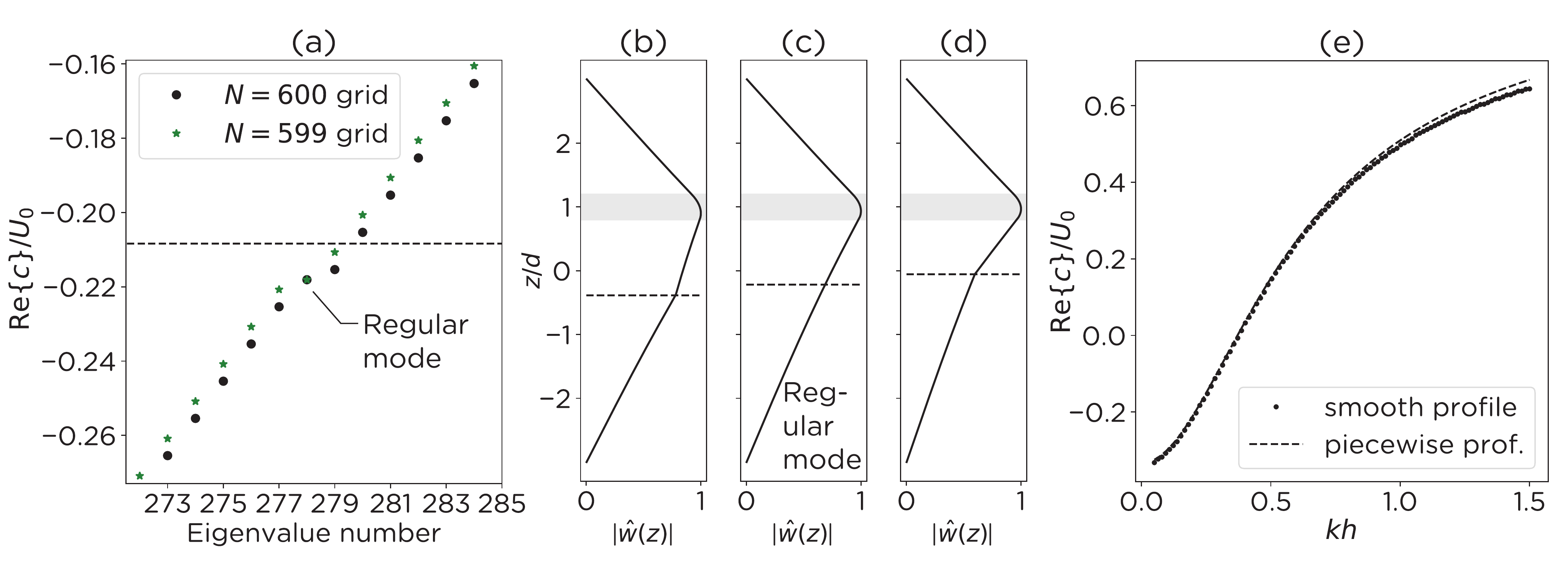} 
\caption{The identification of a stable vorticity wave on the smooth vorticity interface shown in Fig. \ref{f:profiles}(a) for a $kh = 0.2$.  A selection of the sorted distribution of eigenvalues $c = \mathrm{Re}\{c\}$ for two slightly different grids, corresponding to $N = 600$ (black dots) and $N = 599$ (green stars) grid levels, is shown in (a).  The numbering of the eigenvalues on the horizontal axis is arbitrary.  The dashed horizontal line in (a) indicates the phase speed of the piecewise profile.  Selected eigenfunctions with phase speeds less than, equal to, and greater than, the regular mode phase speed are plotted in (b,c,d), respectively.  The horizontal dashed line marks the location of the critical height in each case, and the grey shading the region where $U'' \neq 0$.  The dependence of the regular mode phase speed on dimensionless wavenumber, $kh$, is shown in (e), along with the dispersion relation of the piecewise $U(z)$ as a dashed line.}
\label{f:isolation}
\end{center}
\end{figure*}

Such an analysis has been conducted for the smooth $U(z)$ that has previously been suggested by Baines et al.\cite{bain1996}, and is shown in Fig. \ref{f:profiles}(a).  It is composed of constant shear above and below an interfacial region of thickness $d$, in which $U'' = -(2hd)^{-1} = \mathrm{const.}$, that can be described by the function
\begin{equation} \label{eq:baines_profile}
\frac{U(z)}{U_0} = \Bigg\{ \begin{array}{l@{\quad}l} 1  & \textrm{, $z>h + d$} \\ 
1 - (h + d - z)^2/(4 h d) & \textrm{, $h - d < z < h + d$} \\ 
z/h & \textrm{, $z < h - d$}.\\ \end{array} 
\end{equation}
A plot of a selection of the eigenvalues that result, sorted by phase speed, is shown in Fig. \ref{f:isolation}(a).  These modes correspond to the continuous spectrum, and exhibit eigenfunctions that have singularities located at the critical height (e.g., Figs. \ref{f:isolation}b and \ref{f:isolation}d, where $|\hat{w}(z)|$ is plotted). This can be seen when we change the numerical grid from $N = 600$ grid points to $N=599$, and observe a corresponding change in the real part of the phase speed, $\mathrm{Re}\{c\}$, with the imaginary part being zero (Fig. \ref{f:isolation}a).  There is however, a single exceptional mode that can be seen in Fig. \ref{f:isolation}(a).  This "\emph{regular mode}" exhibits very little change with $N$, and corresponds to an eigenfunction that does not have a discontinuous derivative at the critical height (Fig. \ref{f:isolation}c).  These are desirable properties that result in a more physically reasonable modal structure, as (i) perturbation vorticity is not introduced in a region of the flow that has no background vorticity gradients, i.e., $U'' = 0$, and (ii) the phase speed is not dependent on the numerical grid.  Point (ii) can be confirmed by doubling $N$, and observing only a $1$\% change in the phase speed of the regular mode.  We argue that this regular mode is the smooth profile analogue of a vorticity wave, defined precisely for piecewise-linear $U(z)$, and is a mode of the discrete, rather than the continuous, spectrum.  


A distinction between modes of continuous and discrete spectra can be made based on the behaviour of $\hat{w}(z)$ at $z_c$.  From inspection of Rayleigh's equation (\ref{eq:rayleigh}) where $U'' = 0$, i.e., $(U-c)(\hat{w}'' - k^2\hat{w}) = 0$, it can be seen that any value of $\hat{w}''$ can be chosen at $z_c$ that allows $\hat{w}$ to satisfy the boundary conditions for that particular $c$, and that $c$ can take on a continuum of values.  Such a solution will have a jump in $\hat{w}'$ at $z_c$.  This is the singular continuous spectrum.  The analysis above shows, however, that there is a special, discrete $c$ where no jump is needed to satisfy the boundary conditions.  This mode therefore, belongs to the discrete spectrum, and is what we refer to as the regular mode.  The identification of such modes is more complicated when $U''(z_c) \neq 0$, but can be treated in a similar fashion.\cite{iga2013}

For piecewise-linear profiles, as shown in Fig. \ref{f:profiles}(a) by the dashed line in the inset, we may write the background vorticity gradient as a sum of delta functions, i.e.,
\be 
U''(z) = \sum_{j=1}^N \Delta Q_j \delta(z-z_j) ,
\ee 
where the sum is over all $N$ vorticity interfaces, defined as locations $z_j$, where the background flow vorticity, $U'$, jumps by an amount $\Delta Q_j \equiv U'(z_j^+) - U'(z_j^-)$.  This allows for a significant simplification of Rayleigh's equation (\ref{eq:rayleigh}) to
\be \label{eq:ray_simple}
\hat{w}'' - k^2 \hat{w} = 0 .
\ee
In arriving at this result we have discarded the continuous spectrum arising from cancelling a factor of $U-c$ from the left hand side.  The form in Eq. (\ref{eq:ray_simple}) admits a general solution of
\be \label{eq:w_soln}
\hat{w}(z) = \sum_{j=1}^N A_j G(z,z_j) ,
\ee
where the $A_j$ are complex coefficients to be determined as part of the solution, and the Green's function 
\be \label{eq:greens}
G(z,s) = \frac{i}{S[k(z_u - z_l)]} \: \cdot \left\{ \begin{array}{r@{\quad}l} S[k(z_u - z)] S[k(s - z_l)]  & \textrm{, $z>s$} \\ 
S[k(z_u - s)] S[k(z - z_l)] & \textrm{, $z<s$} \\ \end{array} \right.
\ee
with $S(y)$ denoting the hyperbolic sine function, $\sinh(y)$.  The stability properties of the flow are then found by integrating Eq. (\ref{eq:rayleigh}) across the interface locations, $z_j$, to give a series of jump conditions that allow for the determination of the $A_j$, as well as a dispersion relation of the form $\mathcal{D}(c,k) = 0$.  These jump conditions correspond to the physical requirement of continuity of stress normal to the interface\cite{draz1982}.  By including many interfaces, an approximate solution to smooth profiles could be obtained that is equivalent to a numerical solution method (albeit an inefficient one)\cite{carp2017}.  

We examine a single vorticity interface located at $z=h$ that corresponds to the piecewise limit of the smooth profile in Eq. (\ref{eq:baines_profile}) as $d \to 0$ (Fig. \ref{f:profiles}a).  The vorticity wave phase speed found in this way is given by
\be
c(k) = U_0 + \frac{\Delta Q}{k} \frac{S[k(z_u - h)] S[k(h - z_l)]}{S[k(z_u - z_l)]} ,
\ee
and plotted in Fig. \ref{f:isolation}(e) as a dashed line.  This dispersion relation follows very closely that obtained from the smooth profile when the location of the regular mode is found for each value of the dimensionless wavenumber, $kh$, plotted as points in Fig. \ref{f:isolation}(e).  This lends further support to the regular mode being the smooth profile analogue of the piecewise-linear vorticity wave.  Note also, that we would expect an increasing deviation of the dispersion curves from each other as $kh$ gets large, since then the length scale $d$, characterising the thickness of the vorticity gradient region, becomes important in setting the propagation speed on the smooth vorticity interface.

\begin{figure*}
\begin{center}
\includegraphics[width=0.99\textwidth]{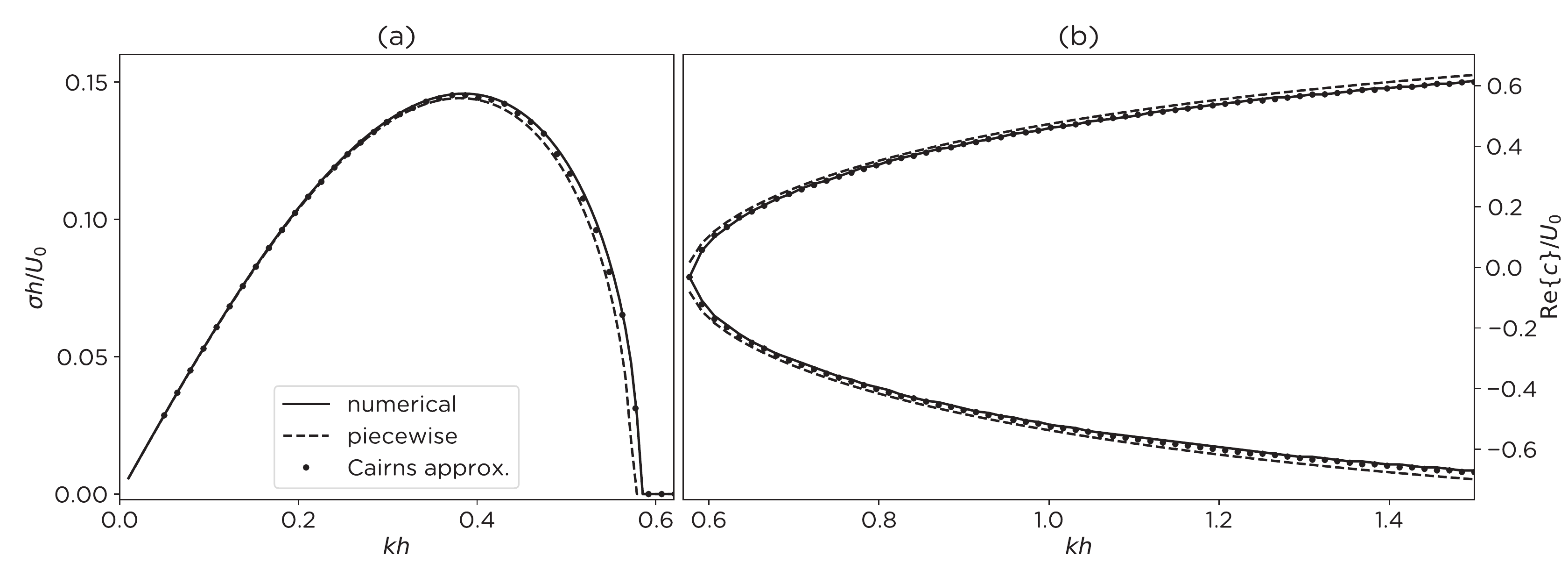} 
\caption{Dimensionless growth rate (a) and phase speed (b) for the shear layers in Fig. \ref{f:profiles}(b).  Solid lines show the numerical results, and dashed lines those for the piecewise-linear profile.  The Cairns' approximation resulting from applying (\ref{eq:cairns}) to the dispersion relation for the smooth vorticity interfaces (as in Fig. \ref{f:isolation}e for the upper interface) is shown by the dots.  }
\label{f:shear_layer}
\end{center}
\end{figure*}

When a second vorticity interface (either smooth or piecewise) with equal and opposite vorticity jump is included at $z=-h$, we introduce an inflection point into the profile, and have a $U(z)$ that is representative of a shear layer (Fig. \ref{f:profiles}b).  This profile supports two vorticity waves that, according to WIT, can interact to cause phase-locking and mutual growth at relatively low values of $kh$, thus producing instability.  This mechanism of shear layer instability through wave interaction is described in many previous studies \cite{holm1962,bain1994,eyal,carp2013,guha2014wave,book}.  Stability results for both the numerical solution of the smooth shear layer, and the analytical solution for the piecewise-linear shear layer, are shown in Fig. \ref{f:shear_layer} in terms of the growth rate, $\sigma \equiv k \mathrm{Im}\{ c \}$, and phase speed, $\mathrm{Re}\{c\}$.  Note that the same method of identifying stable vorticity waves described above, and shown in Fig. \ref{f:isolation}, was used to find the curves in the stable region (Fig. \ref{f:shear_layer}b).  Not surprisingly, the curves follow one another very closely, with differences becoming more apparent for larger $kh$, as expected and discussed for the smooth vorticity interface above.  

Given the isolated dispersion relation of vorticity waves on smooth vorticity interfaces (Fig. \ref{f:isolation}e), it is possible to approximate the growth rate and phase speed characteristics of the unstable shear layer modes directly from WIT.  This can be done by using an approximation due to Cairns\cite{cair1979} (see also Craik\cite{crai1985}), wherein the full dispersion relation of two interacting waves is expanded about the point of crossing of the isolated dispersion curves.  If we denote the appropriate branches of the isolated dispersion curves of the two waves by $c_{1,2}(k)$, then we can write
\be
(c - c_1)(c - c_2) - \mathrm{int}(k) = 0 .
\ee
For general piecewise-linear profiles the interaction term
\be \label{eq:int}
\mathrm{int}(k) = \frac{\mathcal{L}_1|_{c_1} \mathcal{L}_2|_{c_2}}{\frac{\partial \mathcal{D}_1}{\partial c}|_{c_1} \frac{\partial \mathcal{D}_2}{\partial c}|_{c_2} } ,
\ee
where $\mathcal{D}_j(c,k) = 0$ is the isolated dispersion relation of interface $j$.  We have also defined $\mathcal{L}_j$ such that $\mathcal{D}_j \equiv D_j/Dt + \mathcal{L}_j$, with $D_j/Dt \equiv \partial/\partial t + U(z_j)\partial/\partial x$ the linearised material derivative of interface $j$.  The $\mathcal{L}_j$ terms represent the relation between the interface displacement, $\eta_j$, and the vortex sheet strength.  It can be found from the jump condition, applied at the interface level at $z_j$ to give
\be
\frac{D_j}{Dt}A_j + \mathcal{L}_j[\eta_j] = 0,
\ee
where the solution is expressed as in Eq. (\ref{eq:w_soln}), with $A_j$ proportional to the vortex sheet strength.  For example, in this formulation a vorticity interface has $\mathcal{L}_j = ik C_j D_j \eta_j/Dt$, whereas a density interface in a layered stratified flow would have $\mathcal{L}_j = k^2 C_j^2 \eta_j$, with $C_j$ the intrinsic wave speed on each interface, defined by $C_j \equiv c_j - U(z_j)$.

Substituting the vorticity wave dispersion relation into (\ref{eq:int}), the interaction term is given by
\be
\mathrm{int}(k) \equiv \epsilon^2 C_1(k) C_2(k) ,
\ee
with $\epsilon \equiv G(z_1,z_2)/G(z_1,z_1) = G(z_2,z_1)/G(z_2,z_2)$.  Using the quadratic formula allows us to write the dispersion relation as
\be \label{eq:cairns}
c = \frac{c_1 + c_2}{2} \pm \Delta^{1/2} \quad \mathrm{with} \quad \Delta \equiv \frac{(c_1 - c_2)^2}{4} + \mathrm{int}(k) .
\ee
This form, which we will refer to as the Cairns approximation, has the following physical interpretation.  Instability is only possible for $\Delta < 0$, when the difference in the isolated phase speeds of the two interacting modes is smaller than the strength of the interaction between them.  Only when the interaction can counter this propagation difference can phase-locking occur.  In addition, the intrinsic phase speeds of the interacting waves must be opposite so that $\mathrm{int}(k) < 0$, which is sufficient to ensure that they may cause mutual growth in one another, once phase-locked\cite{book,carp2013}.  The strength of the interaction also depends on the decay of the vertical velocity between the two wave-supporting interfaces given by the Green's function in (\ref{eq:greens}), and is quantified by $\epsilon$.  

The Cairns approximation may be applied to the smooth shear layer by using the isolated dispersion relation found for the smooth vorticity interface in Fig. \ref{f:isolation}.  In doing so, we will take the location of the smooth interfaces at $z = \pm h$, and the advection of each wave to be $\pm U_0$.  Note that these choices are approximations, and have no rigorous justification, due to the fact that the interfaces occupy a finite thickness with a variable advection speed.  Nonetheless, the results of applying the Cairns approximation to estimate growth rates and phase speeds, shown in Fig. \ref{f:shear_layer}, exhibit better agreement to the numerical results than the piecewise profile.  We conclude that instability in the smooth shear layer can indeed be understood, and quantified, using WIT.

In closing, we would like to remark that the present work is closely related to that of Iga\cite{iga2013}, who identified two singular, stable interacting waves in the smooth hyperbolic tangent shear layer profile.  We see the present work as an extension of Iga\cite{iga2013} that bridges the piecewise and smooth shear layers.  This connection is complicated by the sudden presence of the singular continuous spectrum for smooth profiles.  However, through the identification of the regular mode a distinction from the continuous spectrum is made, and this mode is shown to be the smooth analogue of the vorticity wave present on sharp interfaces in piecewise-linear profiles.  Finally, utilising the Cairns approximation has allowed for the first quantification of the wave interactions present in the smooth shear layer based on the smooth vorticity waves.  This demonstrates that the wave interaction mechanism is applicable to general smooth shear flows.


\begin{acknowledgments}
We wish to acknowledge the support of the Helmholtz Association (PACES II), and the Alexander von Humboldt Foundation fellowship to AG. We also thank Subhajit Kar for help, and two anonymous reviewers for their insightful comments.
\end{acknowledgments}

\nocite{*}
\bibliography{smooth_profiles}

\providecommand{\noopsort}[1]{}\providecommand{\singleletter}[1]{#1}%
\begin{thebibliography}{17}%
\makeatletter
\providecommand \@ifxundefined [1]{%
 \@ifx{#1\undefined}
}%
\providecommand \@ifnum [1]{%
 \ifnum #1\expandafter \@firstoftwo
 \else \expandafter \@secondoftwo
 \fi
}%
\providecommand \@ifx [1]{%
 \ifx #1\expandafter \@firstoftwo
 \else \expandafter \@secondoftwo
 \fi
}%
\providecommand \natexlab [1]{#1}%
\providecommand \enquote  [1]{``#1''}%
\providecommand \bibnamefont  [1]{#1}%
\providecommand \bibfnamefont [1]{#1}%
\providecommand \citenamefont [1]{#1}%
\providecommand \href@noop [0]{\@secondoftwo}%
\providecommand \href [0]{\begingroup \@sanitize@url \@href}%
\providecommand \@href[1]{\@@startlink{#1}\@@href}%
\providecommand \@@href[1]{\endgroup#1\@@endlink}%
\providecommand \@sanitize@url [0]{\catcode `\\12\catcode `\$12\catcode
  `\&12\catcode `\#12\catcode `\^12\catcode `\_12\catcode `\%12\relax}%
\providecommand \@@startlink[1]{}%
\providecommand \@@endlink[0]{}%
\providecommand \url  [0]{\begingroup\@sanitize@url \@url }%
\providecommand \@url [1]{\endgroup\@href {#1}{\urlprefix }}%
\providecommand \urlprefix  [0]{URL }%
\providecommand \Eprint [0]{\href }%
\providecommand \doibase [0]{http://dx.doi.org/}%
\providecommand \selectlanguage [0]{\@gobble}%
\providecommand \bibinfo  [0]{\@secondoftwo}%
\providecommand \bibfield  [0]{\@secondoftwo}%
\providecommand \translation [1]{[#1]}%
\providecommand \BibitemOpen [0]{}%
\providecommand \bibitemStop [0]{}%
\providecommand \bibitemNoStop [0]{.\EOS\space}%
\providecommand \EOS [0]{\spacefactor3000\relax}%
\providecommand \BibitemShut  [1]{\csname bibitem#1\endcsname}%
\let\auto@bib@innerbib\@empty
\bibitem [{\citenamefont {Baines}\ and\ \citenamefont
  {Mitsudera}(1994)}]{bain1994}%
  \BibitemOpen
  \bibfield  {author} {\bibinfo {author} {\bibfnamefont {P.}~\bibnamefont
  {Baines}}\ and\ \bibinfo {author} {\bibfnamefont {H.}~\bibnamefont
  {Mitsudera}},\ }\bibfield  {title} {\enquote {\bibinfo {title} {On the
  mechanism of shear flow instabilities},}\ }\href@noop {} {\bibfield
  {journal} {\bibinfo  {journal} {J. Fluid Mech.}\ }\textbf {\bibinfo {volume}
  {276}},\ \bibinfo {pages} {327--342} (\bibinfo {year} {1994})}\BibitemShut
  {NoStop}%
\bibitem [{\citenamefont {Carpenter}\ \emph {et~al.}(2013)\citenamefont
  {Carpenter}, \citenamefont {Tedford}, \citenamefont {Heifetz},\ and\
  \citenamefont {Lawrence}}]{carp2013}%
  \BibitemOpen
  \bibfield  {author} {\bibinfo {author} {\bibfnamefont {J.~R.}\ \bibnamefont
  {Carpenter}}, \bibinfo {author} {\bibfnamefont {E.~W.}\ \bibnamefont
  {Tedford}}, \bibinfo {author} {\bibfnamefont {E.}~\bibnamefont {Heifetz}}, \
  and\ \bibinfo {author} {\bibfnamefont {G.~A.}\ \bibnamefont {Lawrence}},\
  }\bibfield  {title} {\enquote {\bibinfo {title} {Instability of stratified
  shear flow: review of a physical mechanism based on interacting waves},}\
  }\href@noop {} {\bibfield  {journal} {\bibinfo  {journal} {Appl. Mech. Rev.}\
  }\textbf {\bibinfo {volume} {64}},\ \bibinfo {pages} {060801} (\bibinfo
  {year} {2013})}\BibitemShut {NoStop}%
\bibitem [{\citenamefont {Guha}\ and\ \citenamefont
  {Lawrence}(2014)}]{guha2014wave}%
  \BibitemOpen
  \bibfield  {author} {\bibinfo {author} {\bibfnamefont {A.}~\bibnamefont
  {Guha}}\ and\ \bibinfo {author} {\bibfnamefont {G.~A.}\ \bibnamefont
  {Lawrence}},\ }\bibfield  {title} {\enquote {\bibinfo {title} {A wave
  interaction approach to studying non-modal homogeneous and stratified shear
  instabilities},}\ }\href@noop {} {\bibfield  {journal} {\bibinfo  {journal}
  {J. Fluid Mech.}\ }\textbf {\bibinfo {volume} {755}},\ \bibinfo {pages}
  {336--364} (\bibinfo {year} {2014})}\BibitemShut {NoStop}%
\bibitem [{\citenamefont {Smyth}\ and\ \citenamefont {Carpenter}(2019)}]{book}%
  \BibitemOpen
  \bibfield  {author} {\bibinfo {author} {\bibfnamefont {W.~D.}\ \bibnamefont
  {Smyth}}\ and\ \bibinfo {author} {\bibfnamefont {J.~R.}\ \bibnamefont
  {Carpenter}},\ }\href@noop {} {\emph {\bibinfo {title} {Instability in
  Geophysical Flows}}}\ (\bibinfo  {publisher} {Cambridge University Press},\
  \bibinfo {year} {2019})\BibitemShut {NoStop}%
\bibitem [{\citenamefont {Redekopp}(2001)}]{rede2001}%
  \BibitemOpen
  \bibfield  {author} {\bibinfo {author} {\bibfnamefont {L.}~\bibnamefont
  {Redekopp}},\ }\bibfield  {title} {\enquote {\bibinfo {title} {Elements of
  instability theory for environmental flows},}\ }in\ \href@noop {} {\emph
  {\bibinfo {booktitle} {Environmental Stratified Flows}}}\ (\bibinfo
  {publisher} {Kluwer, Boston},\ \bibinfo {year} {2001})\BibitemShut {NoStop}%
\bibitem [{\citenamefont {Holmboe}(1962)}]{holm1962}%
  \BibitemOpen
  \bibfield  {author} {\bibinfo {author} {\bibfnamefont {J.}~\bibnamefont
  {Holmboe}},\ }\bibfield  {title} {\enquote {\bibinfo {title} {On the behavior
  of symmetric waves in stratified shear layers},}\ }\href@noop {} {\bibfield
  {journal} {\bibinfo  {journal} {Geofys. Publ.}\ }\textbf {\bibinfo {volume}
  {24}},\ \bibinfo {pages} {67--112} (\bibinfo {year} {1962})}\BibitemShut
  {NoStop}%
\bibitem [{\citenamefont {Carpenter}, \citenamefont {Balmforth},\ and\
  \citenamefont {Lawrence}(2010)}]{carp2010b}%
  \BibitemOpen
  \bibfield  {author} {\bibinfo {author} {\bibfnamefont {J.}~\bibnamefont
  {Carpenter}}, \bibinfo {author} {\bibfnamefont {N.}~\bibnamefont
  {Balmforth}}, \ and\ \bibinfo {author} {\bibfnamefont {G.}~\bibnamefont
  {Lawrence}},\ }\bibfield  {title} {\enquote {\bibinfo {title} {Identifying
  unstable modes in stratified shear layers},}\ }\href@noop {} {\bibfield
  {journal} {\bibinfo  {journal} {Phys. Fluids}\ }\textbf {\bibinfo {volume}
  {22}},\ \bibinfo {pages} {054104} (\bibinfo {year} {2010})}\BibitemShut
  {NoStop}%
\bibitem [{\citenamefont {Iga}(2013)}]{iga2013}%
  \BibitemOpen
  \bibfield  {author} {\bibinfo {author} {\bibfnamefont {K.}~\bibnamefont
  {Iga}},\ }\bibfield  {title} {\enquote {\bibinfo {title} {Shear instability
  as a resonance between neutral waves hidden in a shear flow},}\ }\href@noop
  {} {\bibfield  {journal} {\bibinfo  {journal} {J. Fluid Mech.}\ }\textbf
  {\bibinfo {volume} {715}},\ \bibinfo {pages} {452--476} (\bibinfo {year}
  {2013})}\BibitemShut {NoStop}%
\bibitem [{\citenamefont {Heifetz}\ and\ \citenamefont {Methven}(2005)}]{eyal}%
  \BibitemOpen
  \bibfield  {author} {\bibinfo {author} {\bibfnamefont {E.}~\bibnamefont
  {Heifetz}}\ and\ \bibinfo {author} {\bibfnamefont {J.}~\bibnamefont
  {Methven}},\ }\bibfield  {title} {\enquote {\bibinfo {title} {Relating
  optimal growth to counterpropagating {R}ossby waves in shear instability},}\
  }\href@noop {} {\bibfield  {journal} {\bibinfo  {journal} {Phys. Fluids}\
  }\textbf {\bibinfo {volume} {17}},\ \bibinfo {eid} {064107} (\bibinfo {year}
  {2005})}\BibitemShut {NoStop}%
\bibitem [{\citenamefont {Drazin}\ and\ \citenamefont {Reid}(2004)}]{draz1982}%
  \BibitemOpen
  \bibfield  {author} {\bibinfo {author} {\bibfnamefont {P.~G.}\ \bibnamefont
  {Drazin}}\ and\ \bibinfo {author} {\bibfnamefont {W.~H.}\ \bibnamefont
  {Reid}},\ }\href@noop {} {\emph {\bibinfo {title} {{H}ydrodynamic
  {S}tability}}},\ \bibinfo {edition} {2nd}\ ed.\ (\bibinfo  {publisher}
  {Cambridge University Press},\ \bibinfo {year} {2004})\BibitemShut {NoStop}%
\bibitem [{\citenamefont {Schmid}\ and\ \citenamefont
  {Henningson}(2001)}]{schmid2001stability}%
  \BibitemOpen
  \bibfield  {author} {\bibinfo {author} {\bibfnamefont {P.}~\bibnamefont
  {Schmid}}\ and\ \bibinfo {author} {\bibfnamefont {D.}~\bibnamefont
  {Henningson}},\ }\href@noop {} {\emph {\bibinfo {title} {Stability and
  Transition in Shear Flows}}},\ Vol.\ \bibinfo {volume} {142}\ (\bibinfo
  {publisher} {Springer Verlag},\ \bibinfo {year} {2001})\BibitemShut {NoStop}%
\bibitem [{\citenamefont {Hazel}(1972)}]{hazel1972}%
  \BibitemOpen
  \bibfield  {author} {\bibinfo {author} {\bibfnamefont {P.}~\bibnamefont
  {Hazel}},\ }\bibfield  {title} {\enquote {\bibinfo {title} {Numerical studies
  of the stability of inviscid stratified shear flows},}\ }\href@noop {}
  {\bibfield  {journal} {\bibinfo  {journal} {J. Fluid Mech.}\ }\textbf
  {\bibinfo {volume} {51}},\ \bibinfo {pages} {39--61} (\bibinfo {year}
  {1972})}\BibitemShut {NoStop}%
\bibitem [{\citenamefont {Drazin}(2002)}]{drazin2002introduction}%
  \BibitemOpen
  \bibfield  {author} {\bibinfo {author} {\bibfnamefont {P.~G.}\ \bibnamefont
  {Drazin}},\ }\href@noop {} {\emph {\bibinfo {title} {Introduction to
  Hydrodynamic Stability}}},\ Vol.~\bibinfo {volume} {32}\ (\bibinfo
  {publisher} {Cambridge university press},\ \bibinfo {year}
  {2002})\BibitemShut {NoStop}%
\bibitem [{\citenamefont {Baines}, \citenamefont {Majumdar},\ and\
  \citenamefont {Mitsudera}(1996)}]{bain1996}%
  \BibitemOpen
  \bibfield  {author} {\bibinfo {author} {\bibfnamefont {P.~G.}\ \bibnamefont
  {Baines}}, \bibinfo {author} {\bibfnamefont {S.}~\bibnamefont {Majumdar}}, \
  and\ \bibinfo {author} {\bibfnamefont {H.}~\bibnamefont {Mitsudera}},\
  }\bibfield  {title} {\enquote {\bibinfo {title} {The mechanics of the
  {T}ollmein-{S}chlichting wave},}\ }\href@noop {} {\bibfield  {journal}
  {\bibinfo  {journal} {J. Fluid Mech.}\ }\textbf {\bibinfo {volume} {312}},\
  \bibinfo {pages} {107--124} (\bibinfo {year} {1996})}\BibitemShut {NoStop}%
\bibitem [{\citenamefont {Carpenter}, \citenamefont {Guha},\ and\ \citenamefont
  {Heifetz}(2017)}]{carp2017}%
  \BibitemOpen
  \bibfield  {author} {\bibinfo {author} {\bibfnamefont {J.}~\bibnamefont
  {Carpenter}}, \bibinfo {author} {\bibfnamefont {A.}~\bibnamefont {Guha}}, \
  and\ \bibinfo {author} {\bibfnamefont {E.}~\bibnamefont {Heifetz}},\
  }\bibfield  {title} {\enquote {\bibinfo {title} {A physical interpretation of
  the wind-wave instability as interacting waves},}\ }\href@noop {} {\bibfield
  {journal} {\bibinfo  {journal} {J. Phys. Oceanogr.}\ }\textbf {\bibinfo
  {volume} {47}},\ \bibinfo {pages} {1441--1455} (\bibinfo {year}
  {2017})}\BibitemShut {NoStop}%
\bibitem [{\citenamefont {Cairns}(1979)}]{cair1979}%
  \BibitemOpen
  \bibfield  {author} {\bibinfo {author} {\bibfnamefont {R.}~\bibnamefont
  {Cairns}},\ }\bibfield  {title} {\enquote {\bibinfo {title} {The role of
  negative energy waves in some instabilities of parallel flows},}\ }\href@noop
  {} {\bibfield  {journal} {\bibinfo  {journal} {J. Fluid Mech.}\ }\textbf
  {\bibinfo {volume} {92}},\ \bibinfo {pages} {1} (\bibinfo {year}
  {1979})}\BibitemShut {NoStop}%
\bibitem [{\citenamefont {Craik}(1985)}]{crai1985}%
  \BibitemOpen
  \bibfield  {author} {\bibinfo {author} {\bibfnamefont {A.}~\bibnamefont
  {Craik}},\ }\href@noop {} {\emph {\bibinfo {title} {Wave Interactions and
  Fluid Flows}}}\ (\bibinfo  {publisher} {Cambridge University Press},\
  \bibinfo {year} {1985})\BibitemShut {NoStop}%
\end{thebibliography}%

\end{document}